# Carrier-envelope phase of a light bullet

Elizaveta Zaloznaya[1,2,3], Victor Kompanets[1], Alexander Savvin[3], Alexander Dormidonov[1,3], Sergey Chekalin[1] and Valery Kandidov[1,2]

[1] Institute of Spectroscopy, Russian Academy of Sciences, 108840, Troitsk, Moscow, Russia
[2] Lomonosov Moscow State University, Faculty of Physics, 119991, Moscow, Russia
[3] Dukhov Automatics Research Institute, 127055, Moscow, Russia

E-mail: ed.zaloznaya@physics.msu.ru

**Abstract**

A light bullet is an extremely compressed in space and time wave packet of a few optical cycles, which is formed in the bulk transparent dielectric during femtosecond filamentation under anomalous group velocity dispersion. The effect of a carrier-envelope phase on the dynamics of the light bullet was demonstrated for the first time. The carrier-envelope phase change during a light bullet propagation causes synchronous oscillations of its spatial, temporal and energy parameters with the period decreasing with increasing carrier wavelength. The oscillation period of parameters of a near-single cycle light bullet with broadband frequency-angular spectrum can be described by an analytical estimate written for a Gaussian pulse with a harmonic carrier wave. When analyzing the structure of color centers and induced plasma channels in fluorides, it was experimentally found that light bullet parameters oscillations lead to a periodic change in its nonlinear optical interaction with the dielectric.

Keywords: filamentation, light bullet, carrier-envelope phase, anomalous group velocity dispersion

## 1. Introduction

The development of ultrafast optics opens up new vistas in the sphere of ultrafast metrology the characteristic time of which is determined not by the pulse envelope, but by the period of light field oscillation [1–2]. Ultrafast metrology includes study of inner-shell electronic transitions, charge-transfer in biological molecules and upon chemical processes as well as other issues of electron-optical interaction. These events proceeds on a time scale of the light oscillation period and are incredibly sensitive to absolute phase of a few-cycle light wave. Evolution of an atomic dipole moment, dynamics of light-field-induced electron tunneling, electron energy spectrum - depends not only on envelope amplitude, but also on carrier wave phase in the case of exposure by a few-cycle optical pulse [3–7]. Moreover, this dependence becomes more significant with a reduction in number of optical cycles in a wave packet down to a single cycle.

Filamentation of femtosecond pulses in transparent dielectric under anomalous group velocity dispersion (AGVD) condition causes compression of a wave packet both in time and in space. As a result, an intense few-cycle light bullet (LB) with high electric field localization forms [8-10]. Spatiotemporal light field distribution in a light bullet is dictated by joint manifestation of Kerr and plasma nonlinearities and therefore is rather sophisticated, as well as frequency-angular spectrum, including high harmonics and broadband supercontinuum [10–12]. One estimates the radius of high intensity LB kernel as two-three carrier wavelengths, its duration — as less than two optical cycles [13–15]. Parameters of an LB have been quantitatively determined in [16] by analyzing the distribution of electric field strength simulated with the use of unidirectional pulse propagation equation (UPPE) [17].

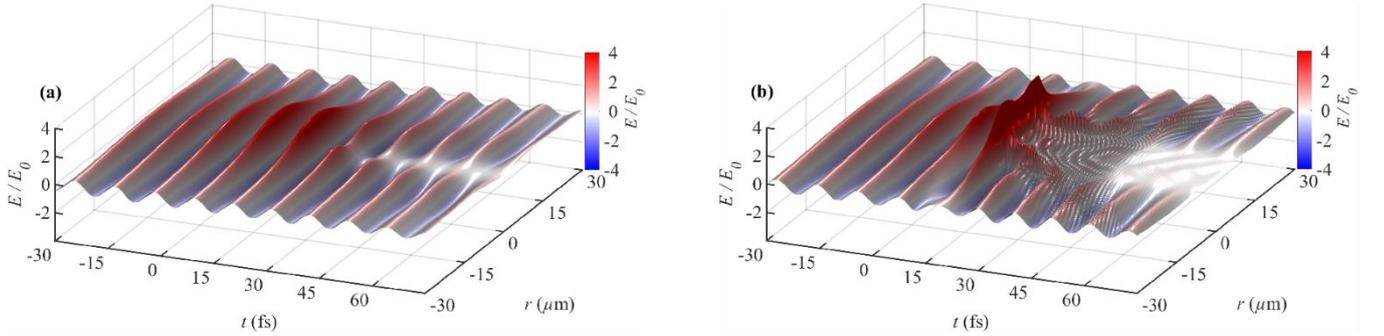

**Figure 1.** Characteristic spatiotemporal distribution of an electric field strength $E(r, t)$ under light bullet formation in a pulse at $\lambda_0$ = 3350 nm for (a) $z$ = 1.6 mm, (b) $z$ = 2.0 mm in LiF. Initial pulse duration $2\tau_0$ = 120 fs, energy $W_0$ = 18 μJ. Pulse propagates in a coordinate frame moving from right to left with the group velocity $V_{gr}(\lambda_0)$.

Absolute phase of the light carrier crucially affects processes accompanying femtosecond filamentation in transparent dielectric. The effect of initial carrier-envelope phase (CEP) on tunneling ionization was found in [18] by recording periodic change in intensity and phase of spectral components during filamentation of an 1800 nm pulse in fused silica. Carrier-envelope phase shift has a dramatic influence on processes of third harmonic and supercontinuum generation during filamentary compression of a pulse down to a few cycles under AGVD [19]. LB formed in a pulse with stable CEP holds it stable [20]. The "breathing" of a few-cycle LB, caused by CEP, was recorded in [21] with laser coloration method. Nevertheless, the influence of CEP on the spatiotemporal and energetic parameters of LB, which determine the efficiency of its nonlinear-optical interaction with a medium, has not been investigated yet.

In this paper, we show synchronous strictly periodic oscillations of the radius, duration and energy during LB propagation in transparent dielectric. The dependence of oscillation period of LB parameters at the carrier wavelength is investigated numerically, on the basis of analysis of electric field strength obtained by solution of unidirectional equation of light bullet propagation in LiF, CaF$_2$, BaF$_2$. The periodic structures induced by LB due to strictly periodic oscillations of its parameters during LB propagation in transparent dielectrics has been experimentally observed as modulation of long lived color centers (CCs) density in LiF and electron concentration of plasma channels in LiF, CaF$_2$, BaF$_2$. By this way a periodic change in the efficiency of LBs nonlinear-optical interaction with the dielectric has been demonstrated. The analytical estimate written for a Gaussian pulse with a harmonic carrier wave can be used to find out the LB oscillation period as a function of wavelength.

## 2. Numerical investigations

### 2.1 Methods

We used unidirectional pulse propagation equation [17] and computer code [22] in numerical investigation of LB formation and propagation in LiF, CaF$_2$ and BaF$_2$. Equation for spectral component of the electric field strength $\tilde{E}(\omega, k_r, z)$ in the nonlinear dispersive medium describes Kerr and plasma nonlinearity, losses by Bremsstrahlung, diffraction and dispersion. Electron concentration in laser plasma was dictated by the process of field ionization, taking into account both multiphoton and tunneling effects [23], as well as process of avalanche ionization due to inelastic collisions of electrons with neutral atoms. In order to study formation and dynamics of an LB during filamentation in fluorides we considered bandwidth-limited collimated Gaussian wave packet

$$E(r,t,z=0) = E_0 exp\left(-\frac{r^2}{2r_0^2} - \frac{t^2}{2\tau_0^2}\right)\cos(\omega_0 t), \quad (1)$$

where $E(r,t,z)$ is an electric field strength, $r_0$, $2\tau_0$ — radius and duration of a wave packet at e$^{-1}$ level of squared field strength; $\omega_0 = 2\pi c_0/\lambda_0$ — frequency at carrier wavelength $\lambda_0$. Initial duration of a multicycle wave packet at the wavelength varying in the range $\lambda_0$ = 2400 – 5700 nm was $2\tau_0$ = 120 fs, initial radius — $r_0$ = 30 μm, peak power $P \approx 1.5 P_{cr}$, where $P_{cr}(\lambda_0)$ — critical power for stationary self-focusing.

### 2.2 Formation of a light bullet with a high-field kernel

Propagation of a wave packet in a bulk Kerr medium under AGVD condition initiates a compression of light field in space and in time due to self-phase modulation caused by cubic nonlinearity (figure 1(a)). An amplitude of electric field strength $E(r,t,z)$ triples during LB formation, therefore, generation of laser plasma takes place. Defocusing in induced plasma gives rise to aberrational distortions when light field sharply decreases on the beam axis. Moreover, local maximum of field strength becomes displaced from the axis which signifies the formation of a ring structure at the pulse tail because of its divergence in plasma (figure 1(b)). Deformations of a wave packet temporal profile gradually come over whole cross-section plane.

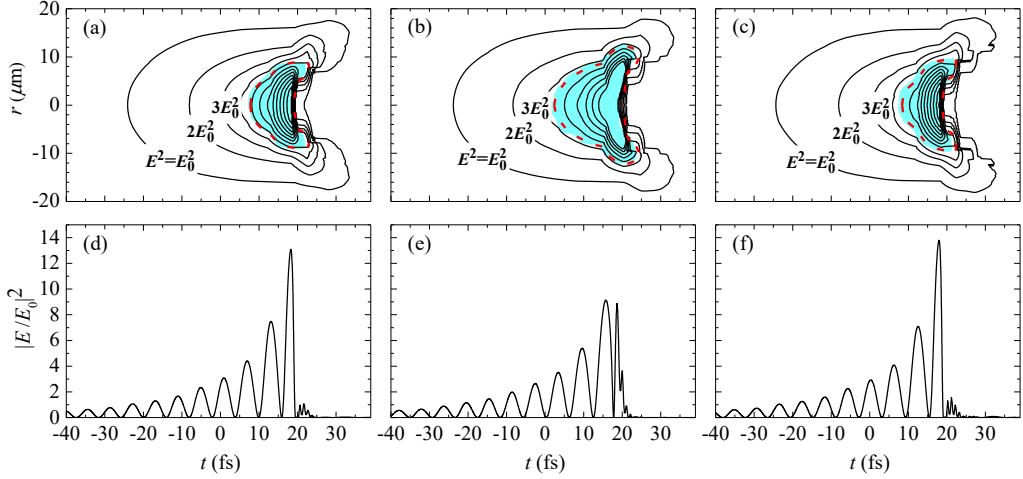

**Figure 2.** (top row) Isolines of squared electric field strength with $E_0^2$ interval. Kernel of a light bullet is a filled region, its border – dashed curve; (bottom row) squared electric field strength on the beam axis $|E(r=0,t,z)|^2$. Shown distributions correspond to nearest maximum and minimum values of $max|E(r=0,t,z)|^2$: (a), (d) $z = 1.780$ mm; (b), (e) $z = 1.795$ mm; (c), (f) $z = 1.810$ mm. Pulse at $\lambda_0 = 3350$ nm propagates from right to left in LiF.

Collection of a few optical cycles with high electric field amplitude is an LB (figure 1(b)). According to [16] we determine parameters of an LB by region of high field localization $S_{HF}$ — kernel of an LB. Squared absolute value of field strength in this region satisfies a condition

$$|E(r,t,z)|^2|_{r,t\epsilon S_{HF}} \geq e^{-1}max_t|E(r=0,t,z)|^2. \quad (2)$$

High-field region has a shape with time axis of symmetry. For a many cycle Gaussian wave packet the border of $S_{HF}$ is an elliptical curve in $(r, t)$ plane. Under compression of a wave packet the high field region monotonically compresses both in space and in time. Its shape remains like initial one up to a light bullet formation. The kernel shape of already formed LB is dramatically distorted (figure 2(top row)). Due to defocusing in induced laser plasma an $S_{HF}$ of LB extremely compresses in time domain and becomes cone-like at the trailing front. Maximum of an electric field shifts toward trailing edge of a wave packet, which testifies the decrease in its group velocity under condition of Kerr nonlinearity.

Maximum radius of $S_{HF}$ determines the local radius $r_e$ of a wave packet. Temporal size of $S_{HF}$ on the beam axis ($r = 0$) shows local duration $2\tau_e$. Along with the local parameters we consider effective radius $r_{eff}$ and effective duration $2\tau_{eff}$ of a wave packet. These quantities take account of electric field distribution in $S_{HF}$:

$$r_{eff} = \left( \frac{\int_{S_{HF}} r^2 |E(r,t,z)|^2 dS}{W_{HF}(z)} \right)^{\frac{1}{2}}, \quad (3a)$$

$$\tau_{eff} = \left( \frac{\int_{S_{HF}} t^2 |E(r,t,z)|^2 dS}{W_{HF}(z)} \right)^{\frac{1}{2}}, \quad (3b)$$

where $W_{HF}(z) = \int_{S_{HF}} |E(r,t,z)|^2 dS$ is an energy in a region $S_{HF}$.

Introduced parameters are the generalization of characteristics of a quasi-harmonic wave packet to an LB consisting of a few cycles. At $z = 0$ mm local $r_e$ and effective $r_{eff}$ radii are the same as radius $r_0$ of an initial wave packet (1), local $2\tau_e$ and effective $2\tau_{eff}$ duration — the same as duration $2\tau_0$.

The considered LB parameters are a lower bound estimate of possible experimental measurements. We have calculated radius and duration of an LB kernel during filamentation of 1800 nm pulse in fused silica with experimental parameters from [20]. One determined parameters of an LB core by high field region with the boundary at half maximum of squared field strength. Calculated in this way effective radius and duration are half as small as those measured in [20].

### 2.3 Light bullet parameters oscillation

Kernel $S_{HF}$ of already formed propagating LB periodically contracts and expands synchronously in space and in time (figure 2(top row)). Peak value of squared field strength on the beam axis $max|E(r=0,t,z)|^2$ also periodically changes with distance (figure 2(bottom row)). It reaches maximum when $S_{HF}$ contracts down to minimal size and decreases with $S_{HF}$ expansion.

Along with periodic change in shape and sizes of LB kernel, parameters of LB oscillate too. Compression of kernel $S_{HF}$ accompanies by synchronous decrease in duration and radius of LB but by corresponding increase in peak field strength. To the contrary, with kernel expansion the duration and radius of an LB increases, but peak field strength decreases. The result is that duration, radius, peak field strength and energy localized in $S_{HF}$ periodically change with LB propagation (figure 3). Energy of LB kernel $W_{HF}(z)$ oscillates in phase opposition opposition with electric field strength, but in phase with sizes of $S_{HF}$ which determine spatiotemporal volume of $S_{HF}$ region. Mean energy $W_{HF}$ localized in a kernel of an LB is about 10% of the total wave packet energy $W_{pulse}$ (figure 3(b)).

During LB propagation in LiF all its parameters oscillate with the same period $\Delta z$, which does not change over the path length. Any deviation of the oscillation period of any parameter from the value averaged over all temporal, spatial and energy parameters of LB is lower than 3%. With an increase in carrier wavelength $\lambda_0$ the oscillation period $\Delta z$ of LB parameters monotonically decreases (figure 4). The same dependence takes place for LB formed in $CaF_2$ and $BaF_2$.

For the physical interpretation of strictly periodic change in the parameters of a few-cycle LB let us extend to it a concept of carrier-envelope phase of a few-cycle pulse. According to [1] carrier-envelope phase shift in a propagating Gaussian few-cycle pulse causes a periodic change in resulting wave amplitude. Spatiotemporal distribution of a field strength in an LB qualitatively stands out from corresponding distribution in a Gaussian pulse or wave packet at the start of compression. LB spectrum is broadband and together with carrier frequency there are high harmonics and high-frequency components of supercontinuum. The notion of the pulse envelope is formally inapplicable for a few-cycle light bullet. We created this envelope using $max|E(r=0,t,z)|^2$ (figure 5). It allowed us to describe axial light field strength in the LB at two close distances as sine- and cosine-shaped waveforms (figure 5(a)). Thus, the light field on the axis of 3350-nm light bullet has a sine-shaped waveform at $z_1 = 1.795$ mm and cosine-shaped waveform at $z_2 = z_1 + \Delta z/2$. These waveforms correspond to 90° CEP shift upon LB propagation over a distance equal to half oscillation period $\Delta z/2 = 15$ μm. It should be noted that at the beginning of wave packet compression during filamentation its shape is similar to many cycle Gaussian one. In this case there is a negligible CEP impact on the resulting amplitude of a light field (figure 5(b)).

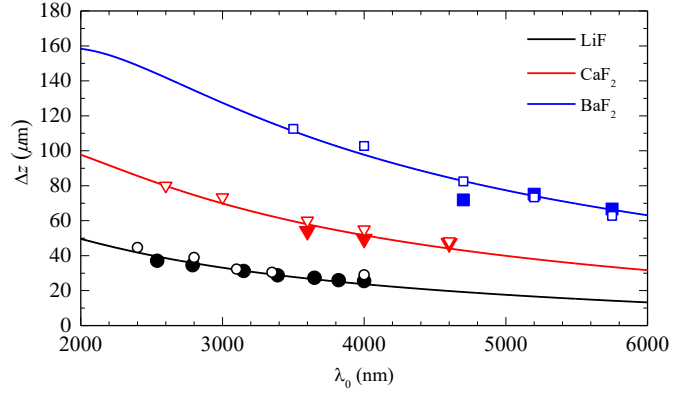

**Figure 4.** Spectral dependence of oscillation period $\Delta z$ of light bullet parameters obtained by computer simulation (empty markers), analytically (4) (solid line), experimentally by measuring color centers density in LiF and electron concentration in $BaF_2$ and $CaF_2$ (filled markers). Experimental errors are in all cases not more than the size of corresponding symbol.

For a pulse with the envelope containing a few cycles of harmonic carrier wave, there is an estimate from [1] of the period $\Delta z$ of the resulting field amplitude oscillation [25]:

$$\Delta z(\lambda_0) = \frac{\lambda_0 V_{gr}(\lambda_0)}{2n(\lambda_0)\left(V_{ph}(\lambda_0) - V_{gr}(\lambda_0)\right)}, \quad (4)$$

where $V_{ph}(\lambda_0)$, $V_{gr}(\lambda_0)$ — tabulated data on phase and group velocity for a wavelength $\lambda_0$. The analytical dependence (4) is in a good agreement with the simulated periods of oscillations of LBs formed in LiF, $BaF_2$, $CaF_2$ (figure 4) despite a key distinction between an LB and an "ideal" pulse. Thus, one can use a simple estimate of period (4) to calculate the oscillation period for parameters of an LB with complex spatiotemporal distribution of a light field and with a broad frequency-angular spectrum.

LB parameters oscillation are due to CEP contrary to periodic changes of field intensity and spectrum associated with interference effects during filamentation under axicon focusing [26,27].

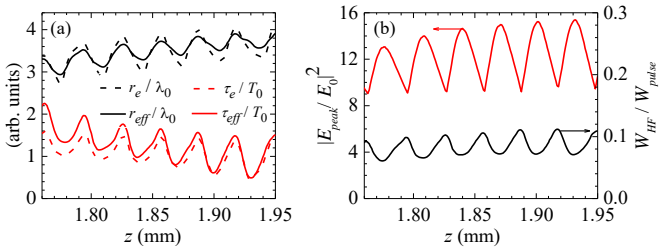

**Figure 3.** Oscillations of light bullet parameters in LiF: (a) local $r_e$ and effective $r_{eff}$ radii, local $2\tau_e$ and effective $2\tau_{eff}$ durations normalized to the wavelength $\lambda_0$ and the period $T_0$ respectively; (b) squared peak strength of electric field on the beam axis $|E_{peak}(r=0)/E_0|^2$ and relative energy $W_{HF}/W_{pulse}$ in the kernel of a light bullet. Pulse wavelength is $\lambda_0 = 3350$ nm.

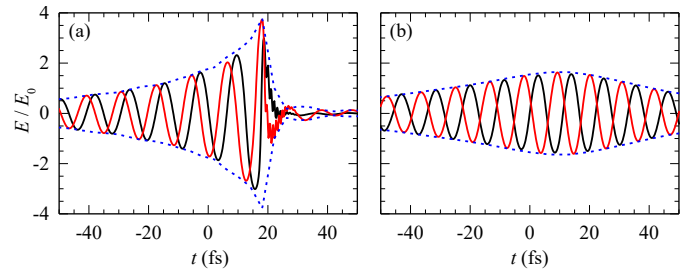

**Figure 5.** Sine- and cosine-shaped electric waves on beam axis: (a) in a light bullet at $z = 1.795$ mm (black line), $z = 1.810$ mm (red line), (b) in a wave packet similar to a Gaussian one at $z = 1.150$ mm (black line), $z = 1.165$ mm (red line); LB envelope approximation (dotted lines). The 3350 nm pulse propagates in LiF from right to left.

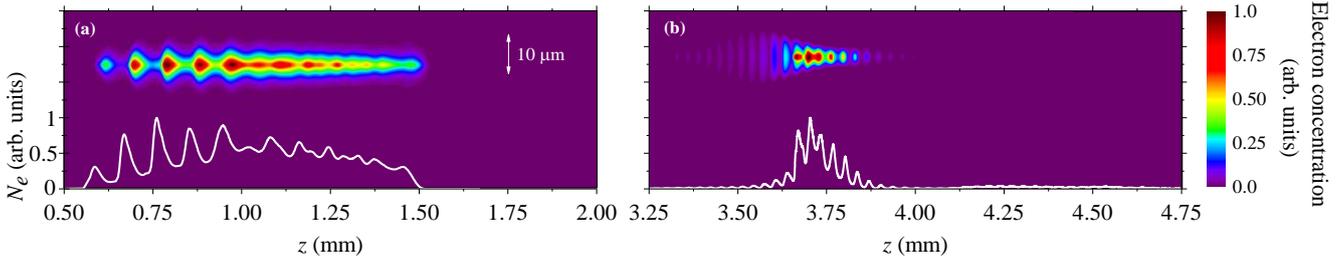

**Figure 6.** Numerically calculated spatial distribution of electron concentration in (a) BaF$_2$ at $\lambda_0$ = 4700 nm (b) LiF at $\lambda_0$ = 3500 nm. On-axis profiles of electron concentration are shown with white curves

The oscillations of the LB parameters lead to the periodic character of its nonlinear interaction with the dielectric. Figure 6 shows numerically calculated plasma channels formed by the LB in BaF$_2$ at $\lambda_0$ = 4700 nm and in LiF at $\lambda_0$ = 3500 nm. The obtained distributions of free electron concentration in plasma channel have a periodic structure corresponding to an oscillating change in the LB electric field strength. The period of change in an electron concentration is $\Delta z$ = 90 μm in figure 6(a) and $\Delta z$ = 31 μm in figure 6(b), which corresponds to oscillation periods shown for the same conditions in figure 4 with empty markers.

## 3. Experimental investigations

In present work the effect of oscillation of LB parameters on its nonlinear interaction was for the first time studied experimentally both by laser coloration method and by high resolution detection of plasma channels glowing that allowed us to investigate different dielectrics.

The experimental layout used for LBs formation was similar to that described in [21,25]. The Mid IR femtosecond pulses at wavelength $\lambda_0$, tunable in the range from 2400 to 5700 nm, which lies in the AGVD region for dielectrics under consideration were generated by the travelling-wave optical parametric amplifier of superfluorescence (TOPAS-C) with the noncollinear difference frequency generator (NDFG). TM polarized 130 fs Mid IR pulses with energy of about 30 μJ were focused by silver-coated concave mirror with a focal length $f$ = 200 mm inside a sample at a distance of several millimeters from its input face. To implement a single-pulse exposure regime the sample was displaced in the direction perpendicular to the laser beam after each shot.

The effect of LB parameters oscillation on its nonlinear interaction with dielectric was studied experimentally by two different methods in a single laser pulse exposure regime, which completely eliminates measurement errors caused by the shot-to-shot irreproducibility of laser pulse parameters. The first one is the laser coloration method [21,25,28] based on the well-known effects of staining of LiF crystals as a result of formation of long-lived structures of luminescent color centers (CCs) to visualize the trace of a laser beam [29]. To analyze the spatial distribution of the luminescence intensity of the recorded CCs filamentary structure that reproduced the density of the laser-induced electronic excitations in the filament we used a monochrome 8Mp CCD camera 8051M (Thorlabs) with 10× NA 0.3 objective with illumination of the induced CCs at the absorption wavelength by a CW laser radiation at 450 nm and detection of luminescence for long time after the writing of these structures. The scattered pump radiation was cut off by an auxiliary yellow-green filter. Some samples of spatial distribution of color centers density in LiF recorded by this way in LiF are shown in figures 7(a)-(b). It should be noted that some evidence for LB parameters oscillations has been experimentally observed in LiF in our preceding works [21,25] by laser coloration method and spectral measurements.

In the second method a structure of plasma channels induced by LB in LiF, CaF$_2$ and BaF$_2$ was registered online for the first time by its self-luminescence straight during LB propagation. These experiments were performed with the use of the same camera with a LOMO 8× NA 0.2 objective. To implement a single laser pulse exposure regime the camera exposure time was set at 0.9 ms. Typical traces of electron concentration spatial distribution along the plasma channel induced by LB obtained by this way in CaF$_2$ are shown in figures 7(c)-(d).

It should be noted that in both cases the length of CCs structures and plasma channels is 300 – 500 μm, which corresponds to LB path length where light field localization is high [19]. Moreover the observed traces of CCs structures and plasma channels clearly demonstrate periodic modulation of signal intensity along the track. It is important that in LiF the oscillation period obtained from CCs structures measurement is equal to one recorded from plasma channels at the same wavelength. In both cases periodic variations of the signal intensity from microstructures induced by multiphoton processes in all dielectrics give evidence of regular change in LB parameters because of the CEP. This is due to the compression of the infrared pulse into a bullet with a duration close to the period of optical oscillations [11,21,28], i.e., indicates the formation of a single-cycle light bullet. Experimentally measured spatial period of this modulation oscillation Δ$z$, caused by CEP decreases with increasing carrier wavelength in a good accordance with calculations (figure 4). It should be noted that all experimental symbols in figure 4 are in the existence domain of near-single cycle LB. Data for $BaF_2$, $CaF_2$ are more scarce than those for LiF due to less domain for short wavelengths and larger filamentation threshold for long wavelengths region.

Results of our experiments show that nonlinear-optical interaction of an LB with dielectric is determined not only by temporal but also by spatial and energetic parameters oscillating during LB propagation.

## 4. Conclusion

As a result of the conducted investigation, we discovered that CEP affects all parameters of a light bullet which is characterized by high localization of a light field both in space

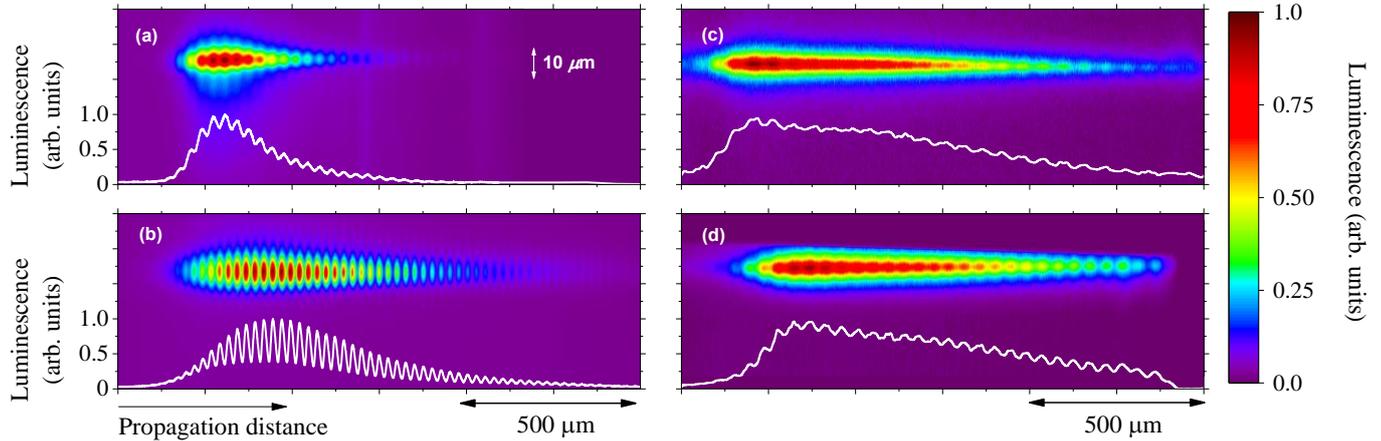

**Figure 7.** Spatial distribution of color centers density in LiF (a) $\lambda_0$ = 2790 nm, (b) $\lambda_0$ = 4000 nm; electron concentration in $CaF_2$ (c) $\lambda_0$ = 3600 nm, (d) $\lambda_0$ = 4600 nm registered in a single-pulse regime. Profiles of color centers density and electron concentration are shown with white curves

and in time. In contrast to the case of a few-cycle Gaussian pulse with a flat wave front [1], upon light bullet propagation, CEP causes periodic variation not only of amplitude of electric field strength but also of its spatiotemporal localization. Oscillation period of a field peak amplitude, duration, radius and light bullet energy decrease with an increase in a carrier wavelength. We found out that the estimate of oscillation period in dependence to wavelength obtained for a Gaussian pulse with harmonic carrier is valid for a light bullet with complex spatiotemporal field distribution and broadband frequency-angular spectrum. Through the analysis of color centers density in LiF and electron concentration in LiF, $CaF_2$, $BaF_2$ induced by a light bullet, we revealed that oscillation of light bullet parameters (caused by carrier-envelope phase) leads to periodic modulation of nonlinear-optical interaction of a light bullet with dielectric. Considering oscillations of spatial and energetic parameters of a light bullet is significant for quantitative analysis of light bullet nonlinear influence upon medium.

## Acknowledgements

Research was supported by Russian Science Foundation (project no. 18-12-00422). E.D. Zaloznaya acknowledges Theoretical Physics and Mathematics Advancement Foundation "BASIS" for the financial support